\documentclass[AMA,STIX1COL]{WileyNJD-v2}

\articletype{Review}%

\received{3 November 2020}
\revised{4  November 2020}
\accepted{5 November 2020}

\begin{document}

% Bibliography and bibfile
\def\aj{AJ}%
          % Astronomical Journal
\def\actaa{Acta Astron.}%
          % Acta Astronomica
\def\araa{ARA\&A}%
          % Annual Review of Astron and Astrophys
\def\apj{ApJ}%
          % Astrophysical Journal
\def\apjl{ApJ}%
          % Astrophysical Journal, Letters
\def\apjs{ApJS}%
          % Astrophysical Journal, Supplement
\def\ao{Appl.~Opt.}%
          % Applied Optics
\def\apss{Ap\&SS}%
          % Astrophysics and Space Science
\def\aap{A\&A}%
          % Astronomy and Astrophysics
\def\aapr{A\&A~Rev.}%
          % Astronomy and Astrophysics Reviews
\def\aaps{A\&AS}%
          % Astronomy and Astrophysics, Supplement
\def\azh{AZh}%
          % Astronomicheskii Zhurnal
\def\baas{BAAS}%
          % Bulletin of the AAS
\def\bac{Bull. astr. Inst. Czechosl.}%
          % Bulletin of the Astronomical Institutes of Czechoslovakia
\def\caa{Chinese Astron. Astrophys.}%
          % Chinese Astronomy and Astrophysics
\def\cjaa{Chinese J. Astron. Astrophys.}%
          % Chinese Journal of Astronomy and Astrophysics
\def\icarus{Icarus}%
          % Icarus
\def\jcap{J. Cosmology Astropart. Phys.}%
          % Journal of Cosmology and Astroparticle Physics
\def\jrasc{JRASC}%
          % Journal of the RAS of Canada
\def\mnras{MNRAS}%
          % Monthly Notices of the RAS
\def\memras{MmRAS}%
          % Memoirs of the RAS
\def\na{New A}%
          % New Astronomy
\def\nar{New A Rev.}%
          % New Astronomy Review
\def\pasa{PASA}%
          % Publications of the Astron. Soc. of Australia
\def\physrep{Phys.~Rep.}%
          % Physics Reports
\def\pra{Phys.~Rev.~A}%
          % Physical Review A: General Physics
\def\prb{Phys.~Rev.~B}%
          % Physical Review B: Solid State
\def\prc{Phys.~Rev.~C}%
          % Physical Review C
\def\prd{Phys.~Rev.~D}%
          % Physical Review D
\def\pre{Phys.~Rev.~E}%
          % Physical Review E
\def\prl{Phys.~Rev.~Lett.}%
          % Physical Review Letters
\def\pasp{PASP}%
          % Publications of the ASP
\def\pasj{PASJ}%
          % Publications of the ASJ
\def\qjras{QJRAS}%
          % Quarterly Journal of the RAS
\def\rmxaa{Rev. Mexicana Astron. Astrofis.}%
          % Revista Mexicana de Astronomia y Astrofisica
\def\skytel{S\&T}%
          % Sky and Telescope
\def\solphys{Sol.~Phys.}%
          % Solar Physics
\def\sovast{Soviet~Ast.}%
          % Soviet Astronomy
\def\ssr{Space~Sci.~Rev.}%
          % Space Science Reviews
\def\zap{ZAp}%
          % Zeitschrift fuer Astrophysik
\def\nat{Nature}%
          % Nature
\def\iaucirc{IAU~Circ.}%
          % IAU Cirulars
\def\aplett{Astrophys.~Lett.}%
          % Astrophysics Letters
\def\apspr{Astrophys.~Space~Phys.~Res.}%
          % Astrophysics Space Physics Research
\def\bain{Bull.~Astron.~Inst.~Netherlands}%
          % Bulletin Astronomical Institute of the Netherlands
\def\fcp{Fund.~Cosmic~Phys.}%
          % Fundamental Cosmic Physics
\def\gca{Geochim.~Cosmochim.~Acta}%
          % Geochimica Cosmochimica Acta
\def\grl{Geophys.~Res.~Lett.}%
          % Geophysics Research Letters
\def\jcp{J.~Chem.~Phys.}%
          % Journal of Chemical Physics
\def\jgr{J.~Geophys.~Res.}%
          % Journal of Geophysics Research
\def\jqsrt{J.~Quant.~Spec.~Radiat.~Transf.}%
          % Journal of Quantitiative Spectroscopy and Radiative Trasfer
\def\memsai{Mem.~Soc.~Astron.~Italiana}%
          % Mem. Societa Astronomica Italiana
\def\nphysa{Nucl.~Phys.~A}%
          % Nuclear Physics A
\def\physrep{Phys.~Rep.}%
          % Physics Reports
\def\physscr{Phys.~Scr}%
          % Physica Scripta
\def\planss{Planet.~Space~Sci.}%
          % Planetary Space Science
\def\procspie{Proc.~SPIE}%
          % Proceedings of the SPIE

\title{The content of astrophysical jets\protect%\thanks{This is an example for title footnote.}
}

\author[1,2]{Gustavo E. Romero*}

%\author[2,3]{Author Two}

%\author[3]{Author Three}

\authormark{Gustavo E. Romero %\textsc{et al}
}

\address[1]{\orgdiv{Instituto Argentino de Radioastronom\'{\i}a (IAR)}, \orgname{CONICET; CICPBA; UNLP}, \orgaddress{\state{Buenos Aires}, \country{Argentina}}}

\address[2]{\orgdiv{Facultad de Ciencias Astron\'omicas y Geof\'{\i}sicas}, \orgname{Universidad Nacional de La Plata}, \orgaddress{\state{Buenos Aires}, \country{Argentina}}}

%\address[3]{\orgdiv{Org Division}, \orgname{Org Name}, \orgaddress{\state{State name}, \country{Country name}}}

\corres{*Gustavo E. Romero,  Instituto Argentino de Radioastronom\'{\i}a (CONICET; CICPBA; UNLP), C.C. No. 5, 1894 Villa Elisa, Argentina. \email{gustavo.esteban.romero@gmail.com}}

%\presentaddress{Instituto Argentino de Radioastronom\'{\i}a (CONICET; CICPBA), C.C. No. 5, 1894 Villa Elisa, Argentina}

\abstract[Summary]{Jets, collimated outflows of particles and fields, are observed in a wide variety of astrophysical systems, including Active Galactic Nuclei of various types, microquasars, gamma-ray bursts, and young stellar objects. Despite intensive efforts along several decades of observational and theoretical research, there are still many uncertainties and open questions related to how jets are produced and what is their composition. In this review I offer an outline of some current views on the content and basic properties of astrophysical jets. }

\keywords{Jets, black holes, magnetic fields, accretion}

\maketitle

\footnotetext{\textbf{Abbreviations:} AGN, Active Galactic Nuclei; ADAF, Advection Dominated Accretion Flows; BH, Black Hole;  DSA, Diffusive Shock Acceleration; EHT, Event Horizon Telescope; FSRQ, Flat Spectrum Radio Quasar; GRMHD, General relativistic magnetohydrodynamics; HD, Hydrodynamics; ISCO, Innermost Stable Circular Orbit; MAD, Magnetically Arrested Disk; MHD, Magnetohydrodynamics; RIAF, Radiatively Inefficient Accretion Flow; SED, Spectral Energy Distribution; SMBH, Super Massive Black Hole; YSO, Young Stellar Object.}

\section{Introduction}\label{sec1}
Astrophysical jets are collimated outflows of particles and fields observed on various scales in several types of both Galactic and extragalactic sources. These sources include Active Galactic Nuclei (AGNs), young stellar objects (YSOs), X-ray binaries (microquasars), and gamma-ray bursts (GRBs). This is a surprising variety of astrophysical systems involving many different scales and numerous physical processes. The properties of jets also span astonishing ranges. For instance, the jets of AGNs can remain well collimated along distances of hundreds of kiloparsecs. In GRBs the jets seem to have initially extreme velocities with Lorentz factors of several hundreds. The extent of jets launched by YSOs can reach several parsecs. Regarding power, jets can go from $10^{32}$ erg s$^{-1}$ in YSOs up to $10^{52}$ erg s$^{-1}$ in GRBs; i.e. they can cover 20 or more orders of magnitude.

The first jet ever observed was discovered by Curtis \cite{Curtis1918} in the nearby galaxy M87, a century ago.  Despite intensive investigation across the entire electromagnetic spectrum, our knowledge of astrophysical jets is still fragmentary and incomplete. One important open issue is related to the composition of these flows. Are they formed by an electron/proton plasma? Or is it an electron/positron fluid? Or a mixture? Are some jets initially Poynting-flux dominated? If so, how do they become matter loaded? What is the origin of the relativistic particles that produce the non-thermal radiation we observe from these jets? And what about the slow, non-relativistic jets we observe in some protostars? The answers to these and other similar questions are closely related to the problem of  how the jets are launched and how they interact with the environment. In this paper I shall discuss these problems, offer some possible answers, and outline prospects for future research. I shall focus mainly on the relativistic jets of AGNs, but I shall occasionally comment on other types of jets.     Because of limitations of space I decided to focus on some general topics avoiding most details. This, I hope, will provide the reader with a general framework to start with. In order to delve into the subject, there are several useful monographs and books\cite{Begelmanetal1984,Blandfordetal2019,Beskin2009,Meier2012,RomeroVila2014}.

%\section{Accretion regimes}\label{sec2}

%\begin{figure}[t]
%\centerline{\includegraphics[width=342pt,height=9pc,draft]{empty}}
%\caption{This is the sample figure caption.\label{fig1}}
%\end{figure}

%\begin{figure*}
%\centerline{\includegraphics[width=342pt,height=9pc,draft]{empty}}
%\caption{This is the sample figure caption.\label{fig2}}
%\end{figure*}

\section{Jet launching}\label{sec2}

The production of astrophysical jets seems to be always associated with the presence of the same basic ingredients: a gravitational potential well, angular momentum, accretion of matter, and magnetic fields. The potential well is provided by a supermassive black hole in the case of AGNS, by a stellar mass black hole or a neutron star in microquasars and GRBs, and by a protostar in YSOs.  The accreted matter with angular momentum comes from the central region of a galaxy in AGNs and from a companion star in microquasars and short gamma-ray bursts; in long GRBs what seems to be accreted is an imploding star; finally in YSOs matter is accreted from the original molecular cloud where the young star is being born.  In all these processes an accretion disk is formed.  Some properties of these systems and the resulting jets are quantified in Table \ref{tab1}.  

\begin{center}
\begin{table}[t]%
\centering
\caption{Properties of jets and central engines of various sources.\label{tab1}}%
\begin{tabular*}{500pt}{@{\extracolsep\fill}lcccc@{\extracolsep\fill}}
\toprule
\textbf{Property} & \textbf{GRBs}  & \textbf{Microquasars}  & \textbf{YSO}  & \textbf{AGN} \\
\midrule
Mass of the accretor [M$_{\odot}$] & $\sim10$ & $\sim10$  & $\sim 1-10$  & $\sim 10^6-10^{9}$ \\ %\tnote{$\dagger$}   \\
Size of the accretor [cm]& $\sim 10^6$ &  $\sim 10^6$  &  $\sim 10^{11}$  &   $\sim 10^{11}-10^{15}$  \\
 Maximum magnetic field [G] &  $\sim 10^{16}$ &  $\sim 10^{7}$  &  $\sim 10^{3}$  &  $\sim 10^{3}-10^{5}$ \\ %\tnote{$\ddagger$}   \\
 Jet power [erg s$^{-1}$]& $\sim 10^{50}-10^{52}$  &$\sim 10^{37}-10^{40}$ & $\sim 10^{32}-10^{36}$  & $\sim 10^{42}-10^{46}$   \\
 Lifetime [yr]&  $\sim10^{-6}$ &  $\sim10^{4}-10^{6}$  & $\sim10^{4}-10^{5}$ & $\sim10^{7}-10^{8}$   \\
 Jet size [pc]& $\sim10$   & $\sim10$   & $\sim 1$  & $\sim 10^5$   \\
\bottomrule
\end{tabular*}
\begin{tablenotes}
%\item Source: Example for table source text.
%\item[$\dagger$] Example for a first table footnote.
%\item[$\ddagger$] Example for a second table footnote.
\end{tablenotes}
\end{table}
\end{center}

The structure and spectrum of the disks depend on the accretion rate $\dot{M}$ . Three basic kinds of accretion
regimes can be distinguished depending on the relation of the actual accretion rate to the critical rate defined as,
\begin{eqnarray}
  M_{\rm crit}= \frac{L_{\rm Edd}}{c^2}\approx 1.4 \times 10^{17} \frac{M}{M_{\odot}}\;\; \textrm{g\;s}^{-1},
\end{eqnarray}
where
\begin{eqnarray}
  L_{\rm Edd}= \frac{4\pi G M m_p c}{\sigma_{\rm T}}\approx 1.3 \times 10^{38} \frac{M}{M_{\odot}}\;\; \textrm{erg\;s}^{-1},
\end{eqnarray}
is the Eddington luminosity\footnote{This is the luminosity necessary to stop spherical accretion by the radiation pressure generated by the accreted thermal gas.}, $M$ is the mass of the accreting object, and the remaining symbols have their usual meaning. At sub-critical rates $\dot{M} < \dot{M}_{\rm crit}$ the disk is geometrically thin and optically thick; it can be described using the standard model developed by Shakura and Sunyaev\cite{ShakuraSunyaev1973}. At very low accretion rates, $\dot{M} << \dot{M}_{\rm crit}$, optically-thin advection-dominated regimes with very hot flows are possible\cite{NarayanYi1995,YuanNarayan2014}. Finally, if the regime is super-critical,  $\dot{M} >> \dot{M}_{\rm crit}$,
the disk becomes optically-thick, geometrically slim or thick, and advection-dominated\cite{Abramowiczetal1980,BegelmanMeier1982,Wiita1982}.

The magnetic field in the disk is expected to be initially only toroidal. A poloidal magnetic flux can be generated \textit{in situ} through a turbulent dynamo powered by magnetorotational instabilities\cite{ToutPringles1996,Liskaetal2020}. When the disk becomes a magnetically arrested disk (MAD) in its inner region, powerful jets can be launched with Lorentz factor $\Gamma_{\rm jet}\sim10$ in the case of AGNs.

Powerful, magnetically dominated outflows can be driven either from the accretion disk or the black hole magnetosphere\footnote{For YOSs clearly only the first possibility is available.}.
Poloidal magnetic field lines inclined at angles $i > 30$ degrees to the disk rotation axis present an effective
potential that decreases along field lines. An outflow can be launched and driven away by centrifugal forces in such a situation even for very low coronal temperatures\cite{BlandfordPayne1982}.  For angles $i < 30$ degrees, the coronal plasma cannot be freely driven by the centrifugal forces: instead, it must first overcome the effective-potential barrier, which has a maximum far away from the disk. Therefore, a strong thermal or radiative assistance is required to initiate outflows with such geometry. This radiative assistance is naturally expected in super-critical sources where the radiation pressure is dynamically important. In any case, sub-relativistic, matter-loaded winds from the accretion disks are expected in most cases. For YSOs this is the only mechanism operating for launching jets. 

The production of very strong and relativistic outflows requires a large fraction of the gravitational energy of the accreting matter to be converted to Poynting flux. This can only occur in the very central region of the sources. The basic features of the magnetic model for the generation of relativistic jets are  the following\cite{Koideetal2000,Koideetal2002,Meieretal2001,Tchekhovskoy2015,RomeroGutierrez2020}:

\begin{itemize}
 \item Relativistic jets are produced by rapidly rotating BHs fed by magnetized accretion disks.
 \item The magnetic field geometry in the innermost region is dominated by the poloidal component.
\item The ultimate power source is the rotational energy of the black hole.
\item The energy is extracted via magnetic torque as Poynting flux.
\item Jet collimation is due to the external medium outside the MAD zone\footnote{Magnetic self-collimation due to the toroidal component of the field is not effective in relativistic plasma flows due to kink instabilities\cite{Eichler1993}, so confinement by the pressure and inertia of an external medium seems to be quite essential\cite{GlobusLevinson2016}.}.
\item Jet acceleration is via conversion of the electromagnetic energy into bulk kinetic energy.
\end{itemize}

A torsional Alfv\'en wave is generated by the rotational dragging of the poloidal field lines in the ergosphere of the black hole.  The field develops a toroidal component producing a traveling perturbation. This wave transports magnetic energy outward, causing the total energy of the plasma near the hole to decrease to negative values. When this plasma with negative energy crosses the horizon, the rotational energy of the black hole decreases. Through this process, the energy of the spinning black hole is extracted by the magnetic field. The result is that while plasma is carried into the hole only (and not ejected), electromagnetic power is evacuated along the rotation axis as a Poynting flux. The magnetic field is tied to the infalling plasma, not to the horizon, as in the simplified split monopole model\cite{BlandfordZnajek1977}. I emphasize that it is the back-reaction of the magnetic field which accelerates the ergospheric plasma to relativistic speeds counter to the hole's rotation endowing the plasma with negative energy. And is the accretion of this negative energy plasma which spins down the hole. Since the energy output is due to the black hole and not to the accreting matter (whose role is to produce and support the poloidal field lines), the power of the jet can exceed the accretion power, i.e. 

\begin{eqnarray}
  P_{\rm jet}=q \dot{M} c^2\;\;\; \textrm{with}\;q>1,
\end{eqnarray}
as inferred for some blazars and GRBs\cite{Ghisellinietal2014}. A detailed calculation leads to the following expression for the jet power\cite{BlandfordZnajek1977,Tchekhovskoyetal2010}:

\begin{eqnarray}
P\sim \frac{G^2}{8 c^3} B^2 M_{\rm BH}^2 \; f(a),  \label{P2}
\end{eqnarray}
where $B$ is the field in the ergosphere, $M_{\rm BH}$ is the black hole mass, and $a$ is the dimensionless spin parameter. The function $f(a)$ tends to $a^2$ for low values of $a$ \cite{Tchekhovskoy2015}, yielding the well-known Blandford-Znajek formula:

\begin{equation}
P_{\rm BZ}\propto B^2 M_{\rm BH}^2 a^2  \label{PBZ1},
\end{equation}
or, in convenient units,
\begin{equation}
 P_{\rm BZ}\approx 10^{46}\,\left(\frac{B}{10^4 \rm G}\right)^2 \left(\frac{M}{10^9 M_\odot}\right)^2a^2\,\,\rm{erg\,s}^{-1}.
\label{PBZ2}
\end{equation}
This equation is accurate for $a\leq 0.5$. For higher spin, the formula under-predicts the true jet power by a factor of $\approx 3$ and the 6th order expansion of $f(a)$ is necessary\cite{Tchekhovskoyetal2010}. In any case, Eq. \ref{PBZ2} shows that a rapidly rotating black hole in a MAD regime can produce a Poynting-flux-dominated jet with an efficiency $q_{\rm BZ}=P_{\rm BZ}/\dot{M}c^2 \gg 1$. 

\begin{figure}
\centerline{\includegraphics[width=0.8\textwidth]{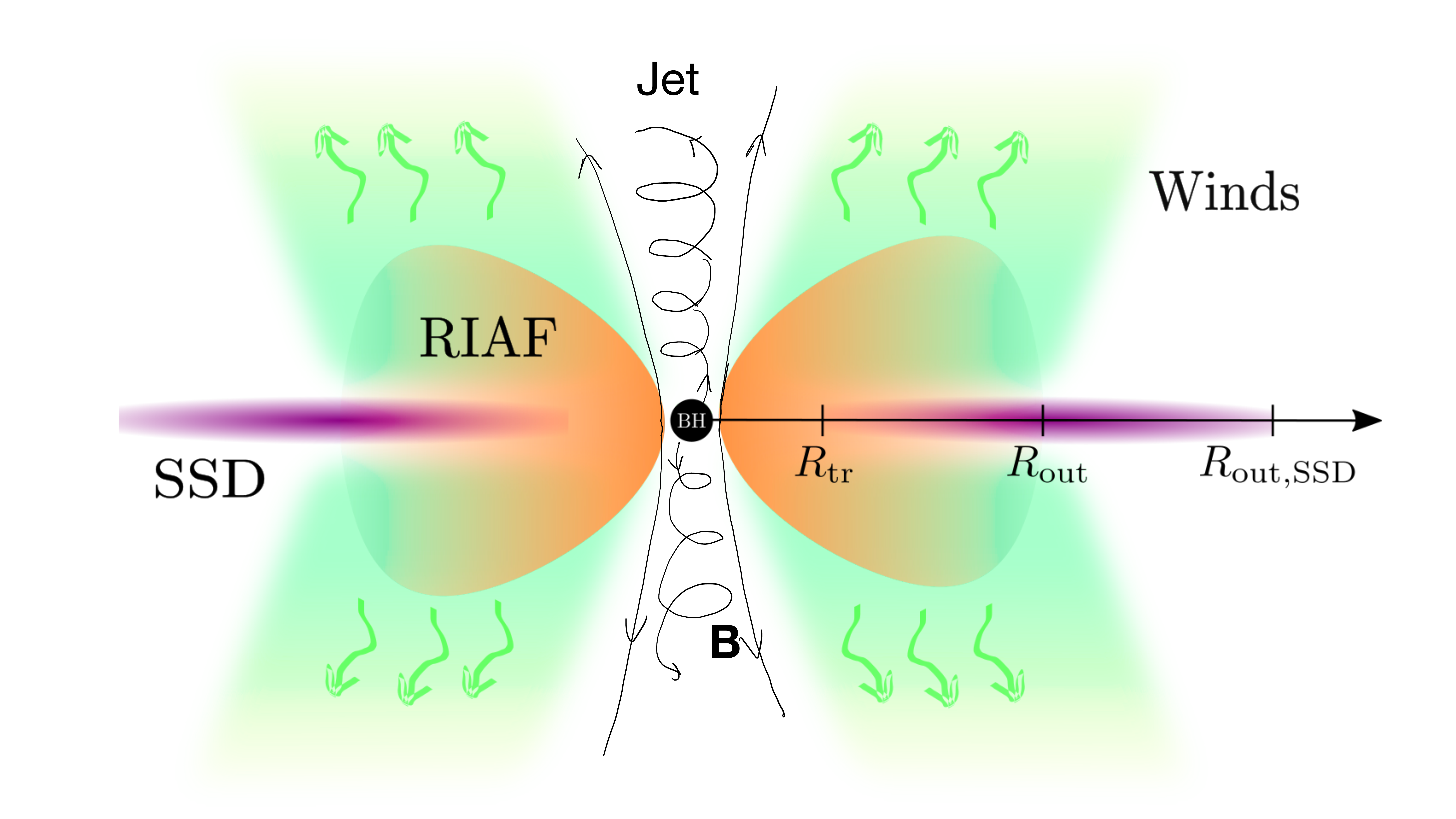}}
\caption{Sketch of the region around a black hole with a radiatively inefficient accretion flow. SSD means the Shakura-Sunyaev thin disk. Not drawn to real scale. Adapted from Guti\'errez et al. \cite{Gutierrezetal2021}. \label{fig1}}
\end{figure}

This Poynting outflow is created in a region where the magnetic field is $\sim 10^4$ G in AGNs (for other types of sources see Table \ref{tab1}). In such sources the radius of the jet is of the order of the Schwarzschild radius of the black hole, i.e. $\sim 10^{15}$ cm. This is much longer than the gyroradius of relativistic particles:
\begin{eqnarray}
  r_{\rm L}=\frac{\gamma_p m_p c^2}{eB}\simeq 3\times10^5 \gamma_p \left( \frac{B}{10^3\; \textrm{G}}\right)^{-1} \;\; \textrm{cm},\end{eqnarray}
where $\gamma_p $ is the Lorentz factor of protons. For electrons the gyroradius is $\sim 10^3$ times smaller than for protons. Clearly, charged particles cannot penetrate the outflow from outside: the jet is magnetically shielded.  The region around the black hole where the jet is launched is illustrated in Figure \ref{fig1} for a radiatively inefficient accretion flow. 

Despite the fact that the evacuation funnel around the rotation axis of the black hole is expected to be free of plasma, very-high-resolution images of the central source of the nearby jet in M87 obtained with the Event Horizon Telescope (EHT)\cite{ETH-1-2019} at 230 GH and Very Long Baseline Array data at 43 and 86 GHz\cite{Hada2013} reveal the existence of radiation associated with the jet at distances down to $\sim10$ gravitational radii from the supermassive black hole\cite{Hada2017}. This radiation is produced by relativistic electrons or positrons injected and accelerated somehow very close to the BH. The jet also accelerates from $\approx0.3c$ at 0.5 mas from the black hole to $\approx2.7c$ at $20$ mas (1 mas $\approx 250 \;r_{\rm g}$)\cite{Parketal2019}, showing an efficient conversion of magnetic energy into bulk motion and internal energy of the emitting gas. What is the origin of this matter so close to the black hole? A possible answer is that charged particles are locally created by gamma-ray photon annihilation and neutron decays. These neutral particles must be produced in the surrounding disk\cite{RomeroGutierrez2020,Vilaetal2014,Toma2012,Kimuraetal2014}.

\section{Matter loading from the hot disk}\label{sec4}

The physics of the accretion of matter onto a black hole is complex since the accretion process develops in many different ways depending on the ambient conditions. To characterize the different regimes, it is useful to parametrize the accretion rate as a function of the Eddington rate:
\begin{equation}
	\dot{m} = \dot{M} / \dot{M}_{\rm Edd},
\end{equation}
where $\dot{M}_{\rm Edd}=10L_{\rm Edd}/ c^2$ (assuming an efficiency of $10\%$).

At very high accretion rates $\dot{m}>>1$ the accretion is super-Eddignton and photon-trapping in the disk becomes important\cite{ohsuga2003,ohsuga2005}: the photon diffusion timescale exceeds the accretion timescale, so photons are advected towards the compact object without being able to reach the surface of the disk and escape. Advection of photons attenuates the disk luminosity, which remains of $\sim L_{\rm Edd}$. Powerful winds are expelled from the disk by the intense
radiation pressure:  practically all the accreted matter is ejected\cite{fukue2000,fukue2004}. This regime occurs in GRBs, AGNs undergoing tidal disruption events, and some microquasars such as SS 433. 

At moderate to high accretion rates, $\dot{m} \gtrsim 0.05$, the usual picture of the accretion flow onto supermassive black holes consists of a geometrically thin ($h/r \lesssim 0.1$), optically thick, cold ($T \ll T_{\rm vir}$) disk that cools efficiently, and a geometrically thick ($h/r \gtrsim 0.5$), optically thin, hot ($T \approx T_{\rm vir}$) corona above and below the disk. The corona is thought to be responsible for the hard X-ray tail detected from active galactic nuclei and microquasars\cite{Bisnovatyi-KoganBlinnikov1977,DiMatteoetal1997,Fabianetal2015,VieyroRomero2012}. This emission is produced by hot electrons via inverse Compton up-scattering of UV/X-ray photons coming from the thin disk. A typical example is the well-known microquasar Cygnus X-1. 

At low accretion rates, $\dot{m} < 0.05$, the thin disk does not extend down to the innermost stable circular orbit (ISCO) and it is truncated at longer distances from the hole. The plasma in the inner region is thought to be in the form of a Radiatively Inefficient Accretion Flow (RIAF) \citep{NarayanYi1995, YuanNarayan2014}. A RIAF is similar to a corona; namely, it is a hot, inflated, optically thin plasma. The hot electrons there Compton up-scatter not only photons from the thin disk but also low-energy (IR) photons produced by synchrotron radiation by the same population of hot electrons ($T\sim10^9$ K). The physical properties of the optically thin, hot component in accretion flows, either a corona or a RIAF, are less understood than those of a cold thin disk. In particular, this hot plasma lies in the so-called {\it collisionless regime}, and thus a fraction of the particle population can be far out of thermal equilibrium. The presence of a non-thermal component in the coronal region leads to the emission of high-energy photons and other particles\cite{Romeroetal2010,Vieyroetal2012,Kimuraetal2019,Inoueetal2019}.

\begin{figure}
\centerline{\includegraphics[width=0.8\textwidth]{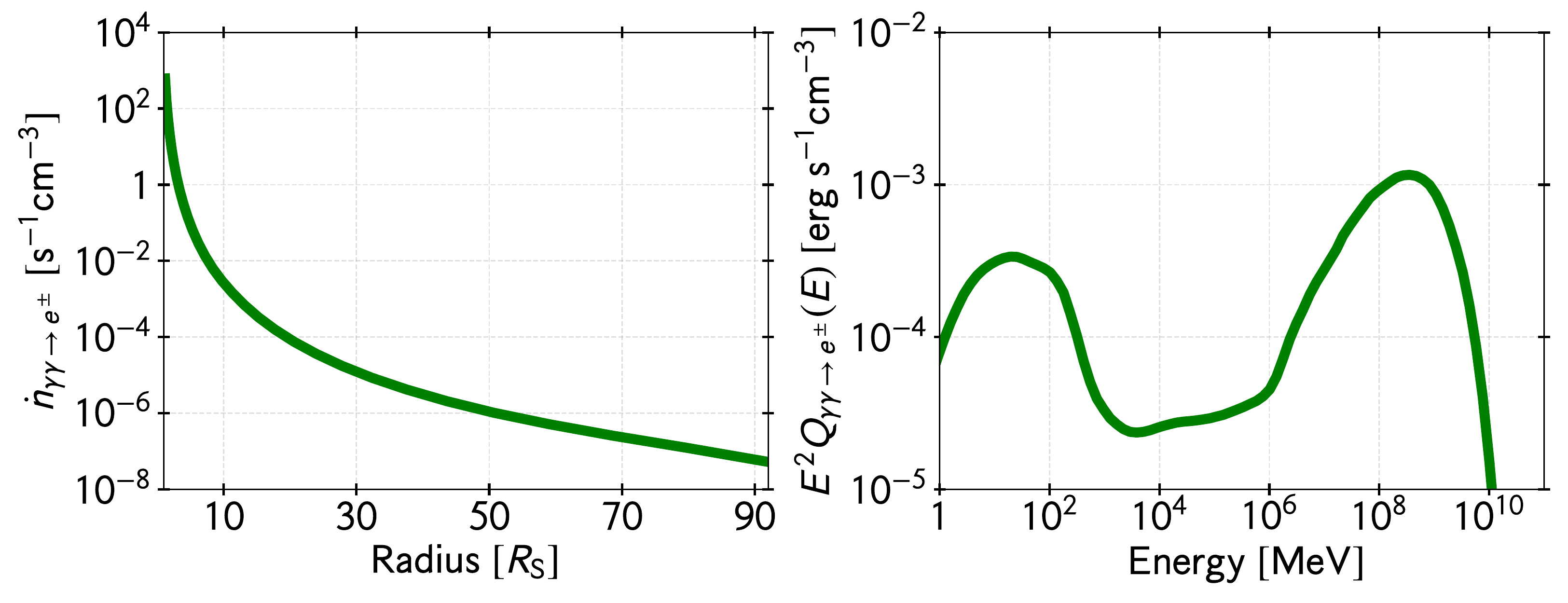}}
\caption{{\it Left panel:} Number of pairs created per unit time and volume by photon annihilation in a hot accretion flow as a function of the radius (distance to the black hole) in units of the Schwarzschild radius. Parameters similar to those of M87 (from Romero and Gutiérrez\cite{RomeroGutierrez2020}). {\it Right panel:} Spectral energy distribution of the same electron/positron pairs in the innermost region of the accretion flow.\label{fig2}}
\end{figure}

\begin{figure}
\centerline{\includegraphics[width=0.5\textwidth]{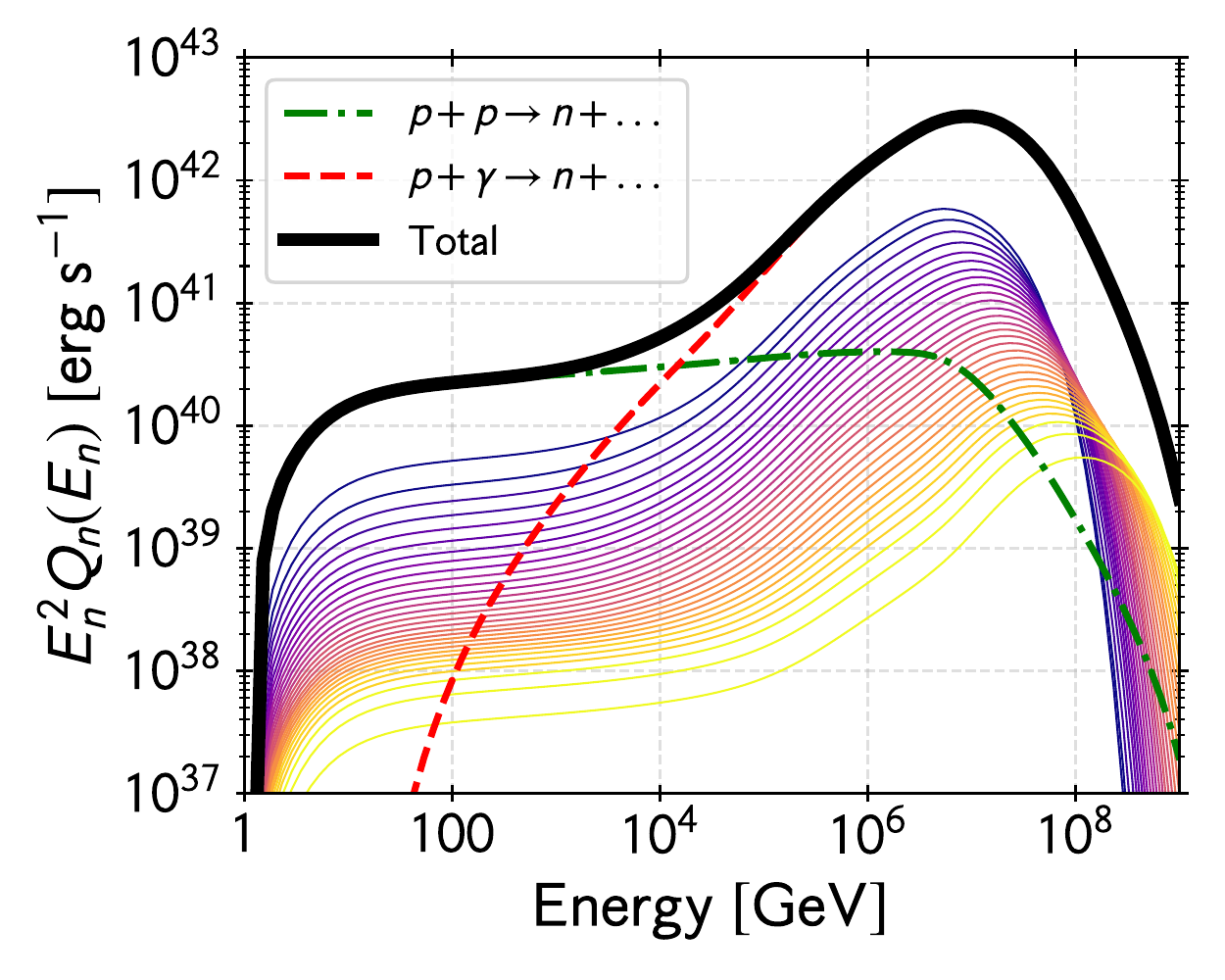}}
\caption{Energy distribution of neutron injection in the same accretion flow of Fig. \ref{fig1}.\label{fig3}}
\end{figure}

Hot accretion flows are more efficient than thin disks in launching and collimating relativistic jets since the conditions discussed in Sect. \ref{sec2} are more easily realized close to the black hole in such regime. The jet is then initiated as a Poynting flux emerging from the BH ergosphere. This flow can be loaded through neutral particles coming from the surrounding hot flow.  More specifically, the particles and involved processes are\cite{RomeroGutierrez2020}:
\begin{itemize}
 \item \textbf{High-energy photons.} They are generated in the hot flow by several processes (most notably proton-synchrotron and proton-photon interactions) and annihilate in the evacuation funnel of the ergospheric jet against background photons.  The annihilation process, $\gamma+\gamma\rightarrow e^- + e^+$, injects pairs at the base of the jet. Direct photon annihilation occurs in two energy bands in the context of hot accretion flows:  two photons with similar energy ($\sim$ MeV) collide producing the low energy peak of the spectral energy distribution, and a high-energy photon annihilates with a low-energy photon (IR, from disk synchrotron radiation). This latter process generates the second peak in the spectrum at higher energies. As an example I show in Figure \ref{fig2} the pair injection spectrum calculated by Romero and Guti\'errez\cite{RomeroGutierrez2020} for a BH similar to that in M87. About $10^3$ pairs per second per cubic centimeter are injected above the black hole in this case. 
 
 \item \textbf{Protons injected by neutron decays.} The neutrons are generated in the hot flow by $pp$ and $p\gamma$ collisions. Energy distributions of neutron injection for both processes in the same conditions as before are shown in Fig. \ref{fig3}. The neutrons can decay inside the jet funnel yielding protons and electrons: $n\rightarrow p + e^- + \bar{\nu_e}$. The protons, in turn, can produce additional pairs by Bethe-Heitler mechanism: $p+\gamma\rightarrow p+e^-+e^+$. More protons can be created by conversion of flying neutrons: $n+\gamma\rightarrow p + \pi^-$. 
\end{itemize}   

Altogether, most protons injected at the base of the jet are the result of neutron decays, although direct neutron conversion into protons might represent a significant fraction close to the black hole. The ratio of the total number of pairs to the total number of protons is $R_{e^\pm / p}\sim 10^5$ and the associated energy ratio is  $R_{u_e^\pm / u_p}\sim 10^2$. Most of the jet mass load by neutral particles in the base of the jet seems to be, then, under the form of pairs. The exact ratio, of course, will change from one source to another depending on the acceleration efficiency and other parameters, but the number of pairs will always exceed the number of protons.  

\section{Entrainment and obstacles}\label{sec5}

The production rate of pairs in the funnel is $\sim 10^{48}$ s$^{-1}$ for an accreting BH similar to the one in M87. This injection implies a power of $L_{e^{\pm}}\sim 10^{42}$ erg s$^{-1}$. If there is not further mass loading, the bulk Lorentz factor of the jet should be $\Gamma\sim P_{\rm BZ}/L_{e^{\pm}}$. If the spin of the black hole is $a\sim 1$, $B\sim 10^4$ G, and $M_{\rm BH}\sim 10^9$ $M_{\odot}$, we have $P_{\rm BZ}\sim 10^{46}$ erg s$^{-1}$ from Eq. (\ref{PBZ2}). Then, $\Gamma\sim 10^4$. This value is far from the typical value $\Gamma\sim10$ inferred from superluminal motions in AGNs. This discrepancy is pointing out to the fact that somehow more baryonic matter should be injected close to the base of the jet by some additional mechanism.

One possibility is the accretion of non-dipolar large scale magnetic fields onto the central object. Numerical simulations show that such accretion creates instabilities that mix disk matter with the outflow charging and slowing down the jet\cite{McKinneyBlandford2009}. The non-dipolar fields are the results of turbulence in the accreting flow. A same source might perhaps switch from dipolar to quadrupolar field accretion, so the jet would decelerate. Then, when the conditions are propitious again, the jet might accelerate and collide with the older outflow forming shocks. These shocks might be responsible for some of the moving knots observed in several sources. 

Baryons can be also supplied by interactions of the jet with different types of obstacles found in its way out. Such obstacles include clouds of the broad line region\cite{Araudoetal2010,delPalacio2019}, clumps in a wind\cite{Araudoetal2009}, or stars\cite{Komissarov1994,Barkovetal2010,BoschRamonetal2012,Araudoetal2013}. In microquasars clumps and interactions with the stellar wind provided by the donor star are likely the main source of protons for the jet\cite{Romeroetal2003,Owockietal2009,Peruchoetal2010,PeruchoValenti2012}.

In the case of the jets of microquasars, where the accretion rate is higher than in most AGNs, baryon load can be important even close to the base of the jet. As a consequence, some jets in these objects become sub-relativistic and dark transporting energy to large distances without being detected unless they find some obstacle that allows part of the kinetic energy to be reconverted into internal energy and then radiated away\cite{Fabrika2004,Galloetal2005,Heinz2006}.

On the other extreme of the power range, the duration distribution of long gamma-ray bursts reveals a plateau at durations shorter than $\sim20$ s (in the observer frame) and a power-law decline at longer durations. Such a plateau arises naturally in the context of the collapsar model because the engine has to operate long enough to push the jet out of the stellar envelope, so the  observed duration of the burst is the difference between the engine's operation time and the jet breakout time. The breakout time for the jet inferred from the duration distribution is similar to the
breakout time obtained with analytic estimates and numerical simulations (both 2D and 3D) of a hydrodynamic jet ($\sim10$ s for typical parameters). This suggests that jets from collapsars are not Poynting flux dominated\cite{Brombergetal2015}.  The mass in these jets might come initially from neutrino annihilation and, after the breakout, from interactions with the external medium.     

\section{Signatures of matter in the jets}\label{sec6}

The power and actual composition of jets can be investigated through their effects on the external medium and from modeling their broadband spectral energy distribution (SED). Studies of the radio and X-ray lobes of radio galaxies suggest that the proton content in the jets tends to be higher in Fanaroff–Riley I (FRI) radio galaxies than in FR IIs\cite{Crostonetal2018}.  Jet-environment interactions also can inform about the jet power and content in the case of microquasars and YSOs. Microquasars such as Cygnus X-1 or SS 433 produce heavy hydrodynamical jets with a significant contribution from cold protons to the total kinetic energy flux.\cite{Heinz2006,Heinz2008}.

The detection of gamma-ray emission that dominates the spectral energy distribution in many blazars can be also used to probe the nature of the jets. This radiation is usually thought to be the result of inverse Compton up scattering of either synchrotron or thermal photons. If the jet is Poynting flux dominated, the synchrotron component should be stronger than observed. This suggests that the outflow is an electron-proton jet, or at most a pair plasma plus an electron-proton fluid\cite{Celotti2003}. 

An upper limit to the pair content of the initial jet can be imposed considering that it should move through the external UV photon field of the inner accretion disk. A large content of pairs in the relativistic flow would Compton up scatter this radiation field producing a  soft X-ray bump that is not observed. The conclusion is  that pure electron-positron jet models can be excluded as over predicting soft X-ray radiation. Similarly, pure electron-proton jets can also be ruled out because they under predict the soft X-ray luminosity of most blazars\cite{SikoraMadjski2000}. What remains is a scenario where the flow starts being Poynting flux dominated but soon is loaded with pairs and electron-proton cold plasma\cite{SikoraMadjski2000,MadejskiSikora2016,RomeroGutierrez2020}. The scales on which the magnetization of the jet drops below unity depend on the initial matter load and the geometry of the jet, but typically should not exceed $\sim10^3-10^4$ Schwarzschild radii\cite{RomeroGutierrez2020}, at least for most AGNs. 

Further insights on jet composition can be obtained from polarization observations\cite{Wardleetal1998}.  The observed radio polarization is dependent on plasma composition due to Faraday conversion and rotation effects. In particular, the contribution of the electrons is degenerate with that of the positrons and their respective number densities
appear additively in the radiative transfer functions relating to emission and linear polarization. The degeneracy is
broken by the consideration of circular polarization, which cancels out for a symmetric $e^-e^+$ plasma, but 
does not of $ep$ plasma. Observations of circular polarization have been used to infer the presence of pairs in jets, but the interpretation of the results depends on various assumptions if the source is not adequately resolved\cite{Homan2018}. Hence, more accurate determinations should wait to detailed polarimetric imaging with the Event Horizon Telescope of M87\cite{ETH-1-2019}. Among the observational signatures predicted for this source are a slight bilateral asymmetry in the flux along the innermost part of the jet and one independent linear polarization component, and bilateral anti-symmetry in the other component. The clearest observable feature is an enhancement of circular polarization with increasing mangetization and ion content\cite{Anantuaetal2020}.

On large scales, by the other hand, the presence of hydrodynamical plasma is revealed by the development of 
substantial back-flowing cocoons. Instead, magnetic dominance of the jet creates  a ``nose cone'' –
a shaped head, which is rarely observed, and might be produced also by some combinations of heavy jets and ambient medium\cite{Komissarov1999,KomissarovFalle2003}. Back-flowing cocoons have been recently observed in microquasars as well, in accordance with the expected hydrodynamic nature of their jets\cite{Martietal2015,Martietal2017}.

The cumulative evidence tends to support a picture of relativistic jets where the outflow starts as a Poynting flux but soon becomes matter dominated. The jet accelerates along distances of $10^3-10^4$ $R_{\rm Schw}$. The matter is an electron-proton plasma plus a pair content, which can vary according to the manifold conditions of different sources. The inertia is determined by the protons, but radiative properties are mostly due to the leptons. The relativistic jet is likely accompanied by a magneto-centrifugal wind driven by the disk.  Mixing and entrainment of matter from the wind is likely.

\section{Conclusions}\label{sec8}

Despite several decades of intensive research on astrophysical jets, many basic issues remain unclear. With the advent of gamma-ray instruments such as \textit{Fermi}, high-resolution X-ray telescopes as \textit{Chandra}, and sub-millimeter interferometry, a consistent picture is finally emerging. Both disks and rotating black holes can launch jets if poloidal magnetic fields are accreted, as predicted in the 1970s and early 1980s. Disks produce magnetohydrodynamic outflows, that are mostly subrelativistic. Magneto-centrifugal outflows formed by $ep$ gas are present in YSOs, microquasars, and some AGNs. Faster, relativistic jets are launched by the ergosphere of Kerr black holes as a Poynting flux but soon become matter dominated. The total power of some of these jets might even exceed the accretion power, clearly indicating that energy is being extracted from the rotation of the black hole. The kinetic power beyond a fraction of parsec from the black hole is dominated by the inertia of protons, but in general the radiative output is due to leptons through synchrotron radiation and inverse Compton interactions. Hadronic radiative processes, however, can be important for the baryon load of the initial jet and during interactions with various kinds of obstacles. The acceleration of hadrons up to ultra-high energies in vacuum electrostatic gaps close to the black hole remains an open possibility that can trigger cascades and high-energy gamma-ray emission in cases with very little accretion\cite{Levinson2000,LevinsonRieger2011,Levinson-Segev2017}.

The predominance of one type of composition or another can of course change along jets, depending on the specific conditions. Forthcoming observations of nearby jets with the EHT will put this picture to robust tests in the near future.

%\backmatter

\section*{Acknowledgments}

This work was supported by the Argentine agency CONICET (PIP 2014-00338), the National Agency for Scientific and Technological Promotion (PICT 2017-2865) and the Spanish Ministerio de
Ciencia e Innovacion (MICINN) under grant PID2019-105510GBC31 and though the ``Center of Excellence Maria de Maeztu 2020-2023'' award to the ICCUB (CEX2019-000918-M).

\subsection*{Author contributions}

This is a review article prepared wholly by the sole author.
 
\subsection*{Financial disclosure}

None reported.

\subsection*{Conflict of interest}

The author declares no potential conflict of interests.

%\section*{Supporting information}

%The following supporting information is available as part of the online article:

%\appendix

%\section{Section title of first appendix\label{app1}}

%\section{Section title of second appendix\label{app2}}%

%== Figure 4 ==
%% Example for figure inside appendix
%\begin{figure}[t]
%\centerline{\includegraphics[height=10pc,width=78mm,draft]{empty}}
%\caption{This is an example for appendix figure.\label{fig5}}
%\end{figure}

%\nocite{*}% Show all bib entries - both cited and uncited; comment this line to view only cited bib entries;
\bibliography{romerobiblio}%

%\clearpage

%\section*{Author Biography}

%\begin{biography}{\includegraphics[width=66pt,height=86pt,draft]{empty}}{\textbf{Author Name.} This is sample author biography text this is sample author biography text this is sample author biography text this is sample author biography text this is sample author biography text this is sample author biography text this is sample author biography text this is sample author biography text this is sample author biography text this is sample author biography text this is sample author biography text this is sample author biography text this is sample author biography text this is sample author biography text this is sample author biography text this is sample author biography text this is sample author biography text this is sample author biography text this is sample author biography text this is sample author biography text this is sample author biography text.}
%\end{biography}

\end{document}